\documentstyle[12pt]{article}

\newcommand{\be}{\begin{equation}}
\newcommand{\ee}{\end{equation}}

\def\psnormal{\textwidth=16cm\textheight=22cm
          \oddsidemargin=0.5cm\evensidemargin=0cm
          \topmargin=0cm\parindent=1cm}
\psnormal

\begin{document}
\pagestyle{empty}

\hspace{3cm}

\vspace{-2.0cm}
\rightline{{ CERN--TH.6507/92}}
\rightline{{ IEM--FT--57/92}}
\rightline{{ US--FT/5--92}}

\vspace{0.8cm}
\begin{center}
{\bf FITTING THE QUARK AND LEPTON MASSES IN STRING THEORIES}
\vspace{1cm}

J.A. CASAS${}^{*,**}\;$, F. GOMEZ${}^{***}$ and C. MU\~NOZ${}^{*}$
\vspace{1.1cm}

${}^{*}$ CERN, CH--1211 Geneva 23, Switzerland
\vspace{0.5cm}

${}^{**}$ Instituto de Estructura de la Materia (CSIC),\\
Serrano 123, 28006--Madrid, Spain
\vspace{0.5cm}

${}^{***}$ Dept. de Fisica Te\'orica,\\
Universidad de Santiago, 15706--Santiago, Spain
\vspace{0.5cm}

\end{center}

\centerline{\bf Abstract}

\vspace{0.5cm}
\noindent The capability of string theories to reproduce at low energy
the observed pattern of quark and lepton masses and mixing angles
is examined, focusing the attention on orbifold constructions, where
the magnitude of Yukawa couplings
depends on the values of the deformation parameters
which describe the size and shape of the compactified space.
A systematic exploration shows that for $Z_3$, $Z_4$,
$Z_6$--I and possibly $Z_7$ orbifolds a correct fit of the
physical fermion masses is feasible. In this way
the experimental masses, which are low--energy quantities, select
a particular size and shape of the compactified space, which turns
out to be very reasonable (in particular the modulus $T$ defining
the former is $T=O(1)$).
The rest of the $Z_N$ orbifolds are rather hopeless and should be
discarded on the assumption of a minimal $SU(3)\times SU(2)\times
U(1)_Y$ scenario. On the other hand,
due to stringy selection rules, there
is no possibility of fitting the Kobayashi--Maskawa parameters
at the renormalizable level, although it is remarked that
this job might well be done by non--renormalizable couplings.

\vspace{0.5cm}
\begin{flushleft}
{CERN--TH.6507/92} \\
{IEM--FT--57/92} \\
{US--FT/5--/92} \\
{May 1992}
\end{flushleft}
\psnormal
\psnormal

\newpage
\pagestyle{plain}
\pagenumbering{arabic}
\section{Introduction}

One of the most intriguing facts of particle physics is the peculiar
experimental pattern of quark and lepton masses and mixing angles.
In the framework
of the Standard Model these are just initial parameters put by hand
without any possible hint about their origin. Grand Unification theories
(GUTs) impose certain relations between them. For instance, in the minimal
SU(5) model, $m_e=m_d$, $m_\mu=m_s$, $m_\tau=m_b$ at $M_{GUT}$. Only the
third equality is compatible with experiments. This is in fact a major
shortcoming of GUTs (also shared by supersymmetric GUTs) which may
only be bypassed by complicating the Higgs sector in an artificial way.
On the other hand, if Superstring Theories are the fundamental theory
from which the Standard Model is derived as a energy limit, they
should be able to give an answer to this fundamental question. This
is the main motivation of the present work. In this sense a crucial
ingredient to relate theory and observation is the knowledge of the
theoretical Yukawa couplings predicted by Superstrings. Actually, there
are several ways to construct four--dimensional strings, but perhaps
the most complete study of the Yukawa couplings has been carried out
for orbifold compactifications
[1--6], which on the other hand have proved to possess very interesting
properties from the phenomenological point of view [7]. So, we will
focus in this letter on this kind of scenarios. Furthermore, orbifold
Yukawa couplings for twisted matter (see below) present a
very rich range, which is extremely attractive as the geometrical
origin of the observed variety of fermion masses [2,8,9].
We will assume throughout that the effective four--dimensional field
theory has $N=1$ supersymmetry, $SU(3)\times SU(2)\times U(1)_Y$ observable
gauge group and three generations of particles with the correct gauge
representations. (We do not consider a GUT theory to avoid the above
mentioned problems.) All these properties have been obtained in explicit
orbifold constructions [10].
Moreover we will assume a unique generation
of Higgses $\{H_1,H_2\}$ (necessary to get a correct Weinberg
angle [11]) and that all the observable matter is of the twisted
type (as was argued in ref.[9], observable untwisted matter
is not phenomenologically viable).

We have made a study, as systematic as possible, for the complete set
of Abelian $Z_N$ orbifolds, i.e. $Z_3$, $Z_4$, $Z_6$--I, $Z_6$--II,
$Z_7$, $Z_8$--I, $Z_8$--II, $Z_{12}$--I, $Z_{12}$--II. We will not
enter here into the details about the construction of these schemes
(these can be found in refs.[1]). Let us recall, however, that a $Z_N$
orbifold is constructed by dividing $R^6$ by a six--dimensional lattice
$\Lambda$ modded by some $Z_N$ symmetry, called the point group $P$.
The space group $S$ is defined as $S=\Lambda\times P$, i.e. $S=\{
(\gamma,u);\ \gamma\in P,\ u\in\Lambda\}$. A twisted string satisfies
$x(\sigma=2\pi)=gx(\sigma=0)$ as the boundary condition, where $g$ is
an element (more precisely a conjugation class) of the space group
whose point group component is non--trivial. Owing to the boundary
condition, a twisted string is attached to a fixed point (sometimes
to a fixed torus) $f$ of $g$. Roughly speaking, the form of $g$ is
$g=(\theta^k, (1-\theta^k)(f+v))$, where $\theta$ is the generator
of $Z_N$ ($\theta^N=1$) and $v\in\Lambda$. It is said
that the string belongs to the $\theta^k$ sector. For a Yukawa
coupling to be allowed the product of the three relevant space group
elements, say $g_1 g_2 g_3$, must contain the identity. This implies
two important equalities:
\begin{equation}
k_1+k_2+k_3=0\;\;{\rm mod}\ N
\label{pgroup}
\end{equation}
\begin{equation}
(1-\theta^{k_1})(f_1+v_1)+\theta^{k_1}(1-\theta^{k_2})(f_2+v_2)-
(1-\theta^{k_1+k_2})(f_3+v_3)=0,\;\;v_i\in\Lambda
\label{sgroup}
\end{equation}
The first one is the so--called point group selection rule, which
implies that the coupling must be of the $\theta^{k_1}\theta^{k_2}
\theta^{-(k_1+k_2)}$ type. The second one is the so--called space
group selection rule, which can have different characteristics
depending on the orbifold under consideration. Some additional
complications appear when a fixed point $f$ under $\theta^k$ is
not fixed under $\theta$ [12,6]. The space group selection
rules for all the $Z_N$ orbifolds have been classified in refs.[5,6].
Likewise, the expressions for the different Yukawa couplings have been
calculated in refs.[2--6]. Their characteristics are summarized in Table 1.
They contain suppression factors that depend
on the relative positions of the fixed points to which the fields
involved in the coupling are attached (i.e. $f_1, f_2, f_3$), and on
the size and shape of the orbifold\footnote{The size and shape of the
orbifold are
given by the vacuum expectation values (VEVs) of certain fields (moduli)
and, consequently, they are dynamical parameters. It has been shown
[13,14] that supersymmetry breaking effects could determine their actual
values.}.
As mentioned above, this has been
suggested as the possible origin of the observed hierarchy of fermion
masses.
We have explored that possibility in this letter, finding that for
certain schemes it can be successfully realized. In section 2 we
discuss the possibility of getting the correct Kobayashi--Maskawa
parameters at the renormalizable level. This turns out to be out
of reach even for non--prime orbifolds, where the mass matrices
are allowed to be non--diagonal. It is remarked, however, that
this job might well be done by non--renormalizable couplings,
at the same time as they account for the masses of
the first generation (which should come from off--diagonal entries
in the mass matrices). However,
renormalizable couplings should still be responsible for
the masses of the second and third generations. Whether this
is possible or not is studied in section 3. As a first step,
a renormalization group analysis is performed, which presents
(slight) differences from the ordinary GUT one. Then it is shown that
for a reasonable size and shape of the compactified space,
the $Z_3$, $Z_4$, $Z_6$--I and possibly $Z_7$ orbifolds can fit the physical
quark and lepton masses adequately. The rest of the $Z_N$
orbifolds, however, should be discarded under the previous
minimal assumptions.
We present our conclusions in section 4.

\section{Mixing angles and geometrical selection rules}

In order to reproduce the mixing angles and the CP violating phase of
the experimental Kobayashi--Maskawa (KM) matrix the quark mass matrices
must have off--diagonal entries. Consequently, the first question is whether
it is possible or not to get non--diagonal mass matrices. The answer to
this question is intimately related to the space--group selection rule
(see eq.(\ref{sgroup})). For prime orbifolds ($Z_3$, $Z_7$), this is of the
so--called diagonal type. This means that given two fields associated
with two fixed points $f_1, f_2$, they can only couple to a unique
third fixed point $f_3$. Of course the coupling, to be allowed,
must satisfy other requirements, in particular gauge invariance.
On the other hand, the couplings must satisfy the point group selection
rule, eq.(\ref{pgroup}), which is also diagonal and, in addition,
in a $\theta^k$ sector the matter associated with a given fixed point
is not degenerate, i.e. all fields have different gauge quantum numbers.
Consequently, for $Z_3$ and $Z_7$ orbifolds the mass matrices are diagonal.
For instance, $Q_uH_2$ (where $Q_u$ denotes the $(u,d)_L$ doublet)
can only couple to a unique field with the gauge
quantum numbers of $u^c$, although this does not mean that such a
coupling must be present. If this were the whole story we should conclude
that mixing angles cannot be obtained within the $Z_3$ and $Z_7$
frameworks. Fortunately, things are quite different when the gauge group
is spontaneously broken after compactification and, in fact, this is
what happens in all the phenomenologically interesting models so far
constructed [10]\footnote{Several sources for this breakdown have been
explored, namely Fayet--Iliopoulos breaking [10], flat directions,
and gaugino condensation induced breaking [14,15].}. Then there
appear new effective trilinear couplings coming from higher order
operators in which some of the fields get non--vanishing VEVs\footnote{
Another (model--dependent) mechanism for mixings, after the breaking,
is explained in refs.[9,16]}. These couplings have a strong
exponential damping [17], but they are no longer subjected to the
trilinear selection rule (examples of this can be found in ref.[10]).
This leads to a natural ansatz for quark and
lepton mass matrices:
\begin{eqnarray}
M=\left( \begin {array}{ccc}
\epsilon & a & b\\
\tilde a & A & c\\
\tilde b & \tilde c & B
\end {array} \right)
\label{CM}
\end{eqnarray}
where $\epsilon,a,\tilde{a},b,\tilde{b},c,\tilde{c}<<A<<B$ in magnitude,
lower--case letters denoting entries generated by higher order operators.
Here we have assumed that the (1,1) entry is zero at the renormalizable
level. As is known, this is extremely convenient to obtain the Cabibbo
angle in a natural manner (more precisely, $\sin\theta_c\sim\sqrt{m_d/m_s}$).
Notice that in this way the masses of the first generation should also be
caused by higher order operators. It is not difficult to construct explicit
models with this property (see e.g. ref.[18]). On the other hand, the entries
$A,B$ should essentially
be generated by renormalizable couplings since non--renormalizable ones are
too small to fit the second and third generation masses properly. Of course,
one has to require $A,B$ to be the correct ones in order to reproduce
those masses. Whether this is possible or not will be studied in the
next section. The ansatz (\ref{CM}) was obtained in ref.[9] in the context
of the $Z_3$ orbifold. It was shown there that it gives correct KM
parameters and first generation masses
for reasonable values of the off--diagonal entries (in
particular it is highly desirable that $\epsilon=0$). Of
course, the precise values of these have to be calculated in each
particular case, but at least this shows that there is no
incompatibility {\em ab initio} between prime orbifolds and the observed
KM parameters. In some sense eq.(\ref{CM}) (with
$\epsilon=0$) is a "stringy" alternative to the Fritzsch ansatz [19]
\begin{eqnarray}
M=\left( \begin {array}{ccc}
0 & A & 0\\
A & 0 & B\\
0 & B & C
\end {array} \right)
\label{Frit}
\end{eqnarray}
(with $|A|<<|B|<<|C|$), which is the most extensively discussed
form for $u$ and $d$--type quark mass matrices.

Things go in a different way for even orbifolds. The reason is twofold.
First, it is clear from Table 1 that, for an even orbifold, Yukawa
couplings are not necessarily of a unique $\theta^{k_1}\theta^{k_2}
\theta^{k_3}$ type. Second, the space group selection rule for a
given $\theta^{k_1}\theta^{k_2}\theta^{k_3}$ coupling is not,
in general, of the diagonal type [6], i.e. for two given fixed points
$(f_1, f_2)$, $f_3$ is not uniquely selected. These two features
in principle open the possibility of having non--diagonal mass matrices
at the renormalizable level, and this is indeed what happens. However,
we will argue now that the structure of these matrices is still
strongly constrained by the selection rules, so that, as for prime
orbifolds, no realistic prediction for the KM parameters can emerge
at the renormalizable level.

Let us first show that the point group selection rule implies that
any viable form for the quark mass matrices should be built up
with Yukawa couplings of a unique $\theta^{k_1}\theta^{k_2}
\theta^{k_3}$ type. Consider for example the $d$--quark mass matrix and suppose
that $H_1, d^c,s^c,b^c$ correspond to the $\theta^l,\theta^{m_1},
\theta^{m_2},\theta^{m_3}$ sectors respectively. Notice now that
if one row of the mass matrix contains more than one entry different
from zero, say $M_{ij_1}, M_{ij_2}\neq 0$, then the point group
selection rule (\ref{pgroup}) requires $l+m_{j_1}=l+m_{j_2}\ \rightarrow
\ m_{j_1}=m_{j_2}$, otherwise the two $SU(2)$ singlet quarks involved
here could not be coupled to the same quark doublet. Now, it is easy
to apply this rule to check that a mass matrix of the Fritzsch type,
eq.(\ref{Frit}), or of the type of eq.(\ref{CM})
cannot be obtained unless all the Yukawa couplings
involved are of the same $\theta^{k_1}\theta^{k_2}\theta^{k_3}$
class. In fact, it is hardly conceivable a phenomenologically
viable mass matrix which does
not contain rows with more
than one non--vanishing entry involving the three generations, so
this rule is general.

Now, we will show that the space group selection rule induces
an important property in the mass matrix which we call
"box--closing" for short. This property means that if we have a $2\times 2$ box
in the mass matrix with three entries different from zero, then the
fourth entry must also be different from zero, e.g.
\begin{eqnarray}
\left[ \begin {array}{cc}
{} & \times \\
\times & \times
\end {array} \right]\;\rightarrow\;
\left[ \begin {array}{cc}
\times & \times \\
\times & \times
\end {array} \right]
\label{BC}
\end{eqnarray}
To see this, suppose that the three initial entries correspond to the
couplings
\begin{equation}
Q_aH_1q_1^c\;,\;\;\;Q_aH_1q_2^c\;,\;\;\;Q_bH_1q_2^c
\label{tresc}
\end{equation}
Calling $\theta^p$, $\theta^l$ the sectors to which $Q_{a,b}$\ ,\  $H_1$
belong, the space group selection rule (\ref{sgroup}) implies
\begin{equation}
(1-\theta^p)(Q_a+v_1)+\theta^{p}(1-\theta^{l})(H_1+v_2)-
(1-\theta^{l+p})(q^c_1+v_3)=0
\label{sgr1}
\end{equation}
\begin{equation}
(1-\theta^p)(Q_a+\tilde v_1)+\theta^{p}(1-\theta^{l})(H_1+\tilde v_2)-
(1-\theta^{l+p})(q^c_2+v_4)=0
\label{sgr2}
\end{equation}
\begin{equation}
(1-\theta^p)(Q_b+v_5)+\theta^{p}(1-\theta^{l})(H_1+\tilde{\tilde v}_2)-
(1-\theta^{l+p})(q^c_2+\tilde v_4)=0
\label{sgr3}
\end{equation}
where $v_i,\tilde v_i,\tilde{\tilde v}_i\in \Lambda$ and we have denoted,
for simplicity, a field and its corresponding fixed point by the same
symbol. Now, (\ref{sgr1})$+$(\ref{sgr3})$-$(\ref{sgr2}) reads
\begin{equation}
(1-\theta^p)(Q_b+v_1+v_5-\tilde v_1)+
\theta^{p}(1-\theta^{l})(H_1+v_2+\tilde{\tilde v}_2-\tilde v_2)-
(1-\theta^{l+p})(q_1^c+v_3+\tilde v_4-v_4)=0
\label{sgr4}
\end{equation}
which implies that the coupling $Q_bH_1q_1^c$ is also allowed\footnote{
The above analysis is more involved when some of the fixed points
are not invariant under $\theta$. However, after an exhaustive study,
it turns out that the box--closing property holds in all cases.}.
This excludes the possibility of obtaining the Fritzsch matrix
(\ref{Frit}) at the renormalizable level (starting with (\ref{Frit})
and applying the box--closing property four times we fill all the
entries). Also the matrix of eq.(\ref{CM}) with $\epsilon=0$
is not allowed. Again, it is hard to imagine any viable mass matrix
satisfying the box--closing property.

This seems to exclude any possibility of having a reasonable mass
matrix at the renormalizable level. One could still try to get something
similar to the Fritzsch matrix, for example, but with very suppressed
couplings instead of zeros. However, this is hopeless since
the selection rule not only imposes the "closing" of any $2\times 2$
box with three non--vanishing entries, see (\ref{BC}), but it usually
relates the value of the fourth entry to those of the three initial ones.
When this is not so, the corresponding suppression factors for the
would--be zero entries should be very
strong, thus requiring high values for the moduli (see next section)
and making virtually impossible to fit the fermion masses correctly .
In view of these results we have to give up fitting the KM parameters
at the renormalizable level. As was mentioned above, this job can
be realized by the (model--dependent) non--renormalizable operators
for the mass matrix of eq.(\ref{CM}), at the same time as they account
for the first generation masses\footnote{This mechanism has also been
considered in ref.[20] in the context of a flipped string model [21].}.
However, the renormalizable couplings should be able to fit the fermion
masses of the second and third generations, which is still extremely
restrictive. This is what we study in the next section.

\section{Fermion masses}
\subsection{Renormalization group analysis}

It is customary to give the experimental values of fermion masses [22]
(except $m_t$) at $1\ GeV$, see first row of Table 2. On the other
hand, the Yukawa couplings in orbifolds are calculated at the
string scale $M_{Str}=0.527\times g\times 10^{18}\ GeV$ [23], where
$g\simeq 1/\sqrt{2}$ is the corresponding value of the gauge coupling
constant. Thus, in order to compare theory and
experiment a renormalization group (RG) running between these
two scales is necessary. This RG analysis differs from the ordinary
GUT one since, in GUTs, the running of
Yukawa couplings is performed between $M_{GUT}$ and $1\ GeV$, where
$M_{GUT}$ is the scale at which gauge interactions are unified.
On the other hand, in string theories, there are "stringy" (no GUT)
threshold corrections on the value of the gauge coupling constants,
shifting the actual scale at which they are unified. More precisely,
the value of these threshold corrections depends on the VEVs of
some moduli\footnote{
Explicit expressions for the threshold corrections can be
found in ref.[23].} that, in general, are not the ones involved
in the Yukawa couplings (see next subsection) [6]. It has been
shown in ref.[24] that, for appropriate VEVs of these moduli
the gauge couplings still unify at an effective unification
scale $M_X\simeq 10^{16}\ GeV$, as is phenomenologically required [25].
However, the running of the Yukawa
couplings has still to be made from $M_{Str}$. This fact, for example,
modifies (slightly) the traditional relation $m_b/m_\tau$ at low energy
when one sets $m_b=m_\tau$ at tree level, as will be seen shortly.

Let us write, for the sake of definiteness, the Yukawa Lagrangian
for the second and third generations, at a scale $\mu$
\begin{eqnarray}
{\cal L}_{Yuk}\ & = & \ h_c(\mu)Q_cH_2c^c + \ h_s(\mu)Q_cH_1s^c
+\ h_\mu(\mu)L_\mu H_1\mu^c
\nonumber \\
& + & \ h_t(\mu)Q_tH_2t^c + \ h_b(\mu)Q_tH_1b^c
+\ h_\tau(\mu)L_\tau H_1\tau^c
\label{Lyuk}
\end{eqnarray}
where the capital letters denote $SU(2)$ doublets and the
$h$'s are the Yukawa couplings. The physical masses
at $1\ GeV$ are then given by
\begin{equation}
m_\alpha=h_\alpha (1\ GeV)\nu_1\ ,\; m_\beta=h_\beta (1\ GeV)\nu_2
\label{masfis}
\end{equation}
where $\alpha=s,b,\mu,\tau$\ ; $\ \beta=c,t$ and $\nu_{1,2}=
\langle H_{1,2}\rangle$
are subjected to the bound
\begin{equation}
\nu_1^2+\nu_2^2 = 2\left(\frac{M_W}{g_2}\right)^2=(175\ GeV)^2
\label{cotanus}
\end{equation}
Moreover, electroweak symmetry breaking in the context of the minimal
supersymmetric standard model suggests $\nu_2 >\nu_1$ [26].
In order to relate $h_{\alpha,\beta}(1\ GeV)$ to
$h_{\alpha,\beta}(M_{Str})$ we have to make use of the RG equations
for the Yukawa couplings between these two scales (see e.g. [27,28]).
This has to be done
in several steps since the matter content is not the same
at any intermediate scale. In particular, we assume as usual a unique
supersymmetric mass $M_{S}$ for all the supersymmetric partners
of the standard matter. (Allowing for a differentiation
of the various supersymmetric masses does not modify the results
substantially.) Besides this, there are, of course, the ordinary
quark thresholds. Following a standard RG analysis and working
in the usual limit $h_b,h_\tau<<h_t$,
we find, for the first two generations of quarks and the three
generations of leptons, the following expressions
\begin{eqnarray}
h_{u,c}(1\ GeV) & = & F_u\ h_{u,c}(M_{Str}) =
\left(\frac{\alpha_3(1\ GeV)}{\alpha_3(m_c)}\right)^{\frac{4}{9}}
\left(\frac{\alpha_3(m_c)}{\alpha_3(m_b)}\right)^{\frac{12}{25}}
\left(\frac{\alpha_3(m_b)}{\alpha_3(M_Z)}\right)^{\frac{12}{23}}
\nonumber \\
& \times &
\left(\frac{\alpha_3(M_Z)}{\alpha_3(M_{S})}\right)^{\frac{4}{7}}
\left(\frac{\alpha_3(M_{S})}{\alpha_3(M_{Str})}\right)^{\frac{8}{9}}
\left(\frac{\alpha_2(M_Z)}{\alpha_2(M_{S})}\right)^{\frac{3}{8}}
\left(\frac{\alpha_2(M_{S})}{\alpha_2(M_{Str})}\right)^{\frac{-3}{2}}
\nonumber \\
& \times &
\left(\frac{\alpha_1(M_Z)}{\alpha_1(M_{S})}\right)^{\frac{-17}{168}}
\left(\frac{\alpha_1(M_{S})}{\alpha_1(M_{Str})}\right)^{\frac{-13}{198}}
h_{u,c}(M_{Str})
\label{RGh1}
\end{eqnarray}
\begin{equation}
h_{d,s}(1\ GeV)= F_u
\left(\frac{\alpha_1(M_Z)}{\alpha_1(M_{S})}\right)^{\frac{12}{168}}
\left(\frac{\alpha_1(M_{S})}{\alpha_1(M_{Str})}\right)^{\frac{6}{198}}
h_{d,s}(M_{Str})
\label{RGh2}
\end{equation}
\begin{eqnarray}
h_{e,\mu,\tau}(1\ GeV) & = &
\left(\frac{\alpha_2(M_Z)}{\alpha_2(M_{S})}\right)^{\frac{3}{8}}
\left(\frac{\alpha_2(M_{S})}{\alpha_2(M_{Str})}\right)^{\frac{-3}{2}}
\nonumber \\
& \times &
\left(\frac{\alpha_1(M_Z)}{\alpha_1(M_{S})}\right)^{\frac{-45}{168}}
\left(\frac{\alpha_1(M_{S})}{\alpha_1(M_{Str})}\right)^{\frac{-3}{22}}
h_{e,\mu,\tau}(M_{Str})
\label{RGh3}
\end{eqnarray}
where $\alpha_1,\alpha_2,\alpha_3$ are the gauge couplings
of $U(1)_Y$, $SU(2)$ and $SU(3)$ respectively. For $h_t$, $h_b$ the
RG equations are more complicated since the effect of the
top Yukawa interactions are not negligible here (see e.g. [27]).
After some algebra one arrives at
\begin{eqnarray}
h_t(M_Z) & = &
\left(\frac{E_1'(M_Z)}{1+\frac{9}{16\pi^2}
h_t^2(M_{S})F_1'(M_Z)}\right)^{\frac{1}{2}}
\nonumber \\
& \times & \left(\frac{E_1(M_{S})}{1+\frac{6}{16\pi^2}
h_t^2(M_{Str})F_1(M_{S})}\right)^{\frac{1}{2}}
h_t(M_{Str})
\label{RGh4}
\end{eqnarray}
\begin{eqnarray}
h_b(1\ GeV) & = &
\left(\frac{\alpha_3(1\ GeV)}{\alpha_3(m_c)}\right)^{\frac{4}{9}}
\left(\frac{\alpha_3(m_c)}{\alpha_3(m_b)}\right)^{\frac{12}{25}}
\left(\frac{\alpha_3(m_b)}{\alpha_3(M_Z)}\right)^{\frac{12}{23}}
\nonumber \\
& \times & \left(\frac{E_2'(M_Z)}{[1+\frac{9}{16\pi^2}
h_t^2(M_{S})F_1'(M_Z)]^{1/9}}\right)^{\frac{1}{2}}
\nonumber \\
& \times & \left(\frac{E_2(M_{S})}{[1+\frac{6}{16\pi^2}
h_t^2(M_{Str})F_1(M_{S})]^{1/6}}\right)^{\frac{1}{2}}
h_b(M_{Str})
\label{RGh5}
\end{eqnarray}
where
\begin{eqnarray}
E_1(Q) & = & \left(1-3\frac{\alpha_3(M_{Str})}{4\pi}t\right)^{\frac{-16}{9}}
\left(1+\frac{\alpha_2(M_{Str})}{4\pi}t\right)^3
\left(1+\frac{33}{5}\frac{\alpha_1(M_{Str})}{4\pi}t\right)^{\frac{13}{99}}
\nonumber \\
E_1'(Q') & = & \left(1-7\frac{\alpha_3(M_{S})}{4\pi}t'\right)^{\frac{-8}{7}}
\left(1-3\frac{\alpha_2(M_{S})}{4\pi}t'\right)^{\frac{-3}{4}}
\left(1+\frac{42}{10}\frac{\alpha_1(M_{S})}{4\pi}t'\right)^{\frac{17}{84}}
\nonumber \\
F_1(M_{S}) & = & \int_{Q=M_{Str}}^{Q=M_{S}} E_1(Q)dt\;,\;\;\;\;\;\;
F_1'(M_{Z})=\int_{Q'=M_S}^{Q'=M_{Z}} E_1'(Q')dt'
\nonumber \\
E_2(Q) & = & E_1(Q)
\left(1+\frac{33}{5}\frac{\alpha_1(M_{Str})}{4\pi}t\right)^{\frac{-12}{198}}
\nonumber \\
E_2'(Q') & = & E_1'(Q')
\left(1+\frac{42}{10}\frac{\alpha_1(M_{S})}{4\pi}t'\right)^{\frac{-24}{168}}
\label{Es}
\end{eqnarray}
with
\begin{equation}
t=2\log \frac{M_{Str}}{Q}\;,\;\;\;t'=2\log \frac{M_{S}}{Q'}
\label{tes}
\end{equation}
The experimental values of $\alpha_i(M_Z)$ are (see e.g. ref.[25]):
\begin{equation}
\alpha_1(M_Z)=0.016930(80),\;\;
\alpha_2(M_Z)=0.03395(52),\;\;
\alpha_3(M_Z)=0.125(5)
\label{alfexp}
\end{equation}
from which $\alpha_i$ can be obtained at any scale following
a standard RG analysis. It has been shown in ref.[25] that a
correct perturbative unification demands\footnote{It can be easily checked
that with $M_{S}= 10^3\ GeV$ the gauge coupling constants
(\ref{alfexp}) are unified at $M_X\simeq2.3\times 10^{16}\ GeV$ with
$\alpha(M_X)\simeq 0.0393$.}
$M_{S}\sim10^3\ GeV$, which is the value we insert in
eqs.(\ref{RGh1}--\ref{tes}) (variations of $M_{S}$ within the
errors are negligible for our purposes). It is interesting to
calculate the values of $h_c,h_s,h_b,h_t,h_\mu,h_\tau$ at $M_{Str}$
which would give the measured values of the corresponding
fermion masses. We have represented them in Fig.1 as
functions of $\nu_2$. Of course, these values for the $h$'s should
emerge from the theory. Whether this happens or not is studied
in the next subsection.

Let us finally note that the relation $m_b(1\ GeV)/m_\tau(1\ GeV)$
is obtained from (\ref{RGh5}) and (\ref{RGh3}). If one imposes,
following the usual GUT ansatz, $h_b(M_{Str})=h_\tau(M_{Str})$,
an expression similar to the GUT one is obtained, but only after
substituting
$M_{GUT}$ by $M_{Str}$ and taking into account that $\alpha_3(M_{Str})
\neq \alpha_2(M_{Str}) \neq \alpha_1(M_{Str})$. Of course, the numerical
results are not substantially affected.

\subsection{The fits}

The theoretical Yukawa couplings $h(M_{Str})$ to be inserted
in eqs.(\ref{RGh1}--\ref{RGh5}) have been calculated in refs.[2--6].
In order to get a feeling of their main characteristics, let us take
the $Z_4$ orbifold based on an $[SO(4)]^3$ root lattice as a
useful example. The action of $\theta$ on the lattice basis
$(e_1,...,e_6)$ is simply $\theta e_l = e_{l+1}$, $\theta e_{l+1} =
-e_l$ with $l=1,3$ and $\theta e_5 = -e_5$, $\theta e_6 = -e_6$.
Let us call $R_i\equiv |e_i|$ and $\alpha_{ij}
=\cos \theta_{ij}$ with $e_i e_j= R_iR_j \cos \theta_{ij}$.
In the orbifold without deformations $\alpha_{ij}=0$ ($i\neq j$).
However the orbifold can consistently be deformed by a modification
of the values of the so--called deformation parameters\footnote{
The values of the deformation parameters correspond to the VEVs
of certain singlet fields with perturbative flat potential,
called moduli and usually denoted by $T_i$.} [9]. For the $Z_4$
orbifold these are $R_1$, $R_3$, $R_5$, $R_6$, $\alpha_{13}$,
$\alpha_{14}$, $\alpha_{56}$ with $(\alpha_{13}+\alpha_{14})^2\leq 1$.
The sizes of the Yukawa couplings
depend on the values of some of them called
effective deformation parameters [9]. For the $Z_4$ these are
\begin{equation}
R_1,\;R_3,\;\alpha_{13},\;\alpha_{14}
\label{EDP}
\end{equation}
On the other hand, for this orbifold all the twisted couplings are
of the $\theta\theta\theta^2$ type and the selection rule reads
\be
f_1 + f_2 - (1 + \theta) f_3 \in \Lambda ,
\label{Z4selrul}
\ee
where $f_3$ is the $\theta^2$ fixed point. The classification of
the fixed points in terms of the lattice basis can be found in ref.[6].
The value of an allowed Yukawa coupling at $M_{Str}$ turns out to be
\begin {eqnarray}
h_{\theta\theta\theta^2} = gN \sum_{v \in (f_2-f_3+\Lambda)_\bot}
\exp [-\frac {1}{4\pi} \vec{v}^{\top} M \vec{v}]
 = gN\;\;  \vartheta \left[
\begin {array}{c}
\vec{f_{23}} \\
0
\end{array}
\right]
[ 0 , \Omega ] ,
\label{ac41}
\end {eqnarray}
where the subscript $\bot$ denotes projection on the $(e_1,...,e_4)$
$D=4$ space, $f_{23}=f_2-f_3$,
the arrow means the corresponding 4--plet of components
and $\vartheta$ is the Jacobi theta function. Moreover
\begin{eqnarray}
\begin{array}{c}
N = \sqrt{V_{\perp}}\; \frac{1} {2\pi}\; \frac {\Gamma ^2
 (\frac{3}{4})}{\Gamma^2 (\frac
{1}{4})}
\\ \\
M= (-4\pi^2i) \Omega =
\left(
\begin{array}{cccc}
R_1^2 & 0 & R_1R_3\alpha_{13} & R_1R_3\alpha_{14} \\
0     & R_1^2 & -R_1R_3\alpha_{14} & R_1R_3\alpha_{13} \\
R_1R_3\alpha_{13} & -R_1R_3\alpha_{14} & R_3^2 & 0 \\
R_1R_3\alpha_{14} & R_1R_3\alpha_{13} & 0 & R_3^2
\end{array}
\right)
\end{array}
\label{mac41}
\end{eqnarray}
where $V_{\perp}=R_1^2R_3^2(1-\alpha_{13}^2-
\alpha_{14}^2)$ is the volume of the unit cell of the
$(e_1,...,e_4)$ lattice. If $f_3$ is not fixed by $\theta$ the
result for $h_{\theta\theta\theta^2}$ is exactly the same but multiplied
by $\sqrt{2}$. Notice that $h_{\theta\theta\theta^2}$ depends
on the relative positions in the lattice of the relevant fixed
points to which the physical fields are attached. This
information is condensed in $\vec{f_{23}}$. In addition,
$h_{\theta\theta\theta^2}$ depends on the size and shape of
the compactified space, which is reflected in the effective
deformation parameters $(R_1, R_3, \alpha_{13}, \alpha_{14})$
appearing in $\Omega$ and in $V_{\perp}$. Note
that both pieces of information appear in a completely distinguishable way
from each other in eq.(\ref{ac41}). It is also interesting
to say that the number of allowed couplings is 160. The number
of {\em different} Yukawa couplings is 10. These characteristics
are summarized in Table 1 for all the orbifolds.

In order to calculate the value of a specific Yukawa coupling,
say $h_s$ (see eq.(\ref{Lyuk})) we need to know what the
$\theta^k$ sectors are and fixed points to which $Q_c,H_1$ and $s^c$
are associated. Actually, it is an empirical fact that, because of the
huge proliferation of scenarios within a given compactification scheme
($Z_4$ in this case), the observable fields can correspond to
any choice of $\{\theta^k,f\}$ sectors, see e.g. ref.[18]. Consequently,
we will take the freedom to assign the physical fields to $\theta^k$
sectors and fixed points at convenience. Of course, a particular
assignment will only be realized in certain scenarios. It is
interesting to notice, however, that not for all the assignments
are the physical Yukawa couplings (see eq.(\ref{Lyuk})) allowed
from the point group and space group selection rules. In order
to illustrate this, suppose that $H_1$ and $H_2$ belong to the
$\theta$ and $\theta^2$ sectors respectively. Then, writting
the space group selection rule (\ref{Z4selrul}) for the quark couplings
of ${\cal L}_{Yuk}$ (see eq.(\ref{Lyuk})),
one finds after some algebra
\be
(H_2-t^c) - (1 + \theta) (b^c-H_1)=
(H_2-c^c) - (1 + \theta) (s^c-H_1) + \Lambda
\label{assrul}
\ee
where we have denoted the fields and their corresponding fixed points
by the same symbols. Eq.(\ref{assrul}) sets severe restrictions on the
possible assignments and, hence, on the possible correspondences of the
physical Yukawa couplings to the above mentioned 10 different Yukawa
couplings. Similar expressions appear if we initially assign $H_1,H_2$
to other $\theta^k$ sectors.

The final step is to let the effective deformation parameters
vary in order to see whether for some choices of them the theoretical
masses, calculated using
eqs.(\ref{masfis}, \ref{RGh1}--\ref{RGh5}, \ref{ac41}), coincide with
the experimental ones (see also Fig.1). In this fit $\nu_1$ has also
to be considered as a free parameter (within the limits mentioned
in the previous subsection) while $\nu_2$ is given by (\ref{cotanus}).
Of course, a different fit has to be made for each possible assignment.

If a satisfactory fit is found, this means that the corresponding
orbifold scheme (in this case $Z_4$) is compatible with the observed
spectrum of fermion masses, which is highly non--trivial as will
be seen shortly. On the other hand, if no such fit is found the
orbifold scheme should be discarded. Obviously, orbifolds with
a higher number of deformation parameters and different Yukawa
couplings (see Table 1) are in a better position to fit the
experimental masses, but this is not a guarantee. For the particular
case of the $Z_4$ orbifold, we have found that, for most of the
possible assignments, the fits are not satisfactory. However,
provided
\begin {eqnarray}
\vec f_{23}(c) & = & (00\frac{1}{2}\frac{1}{2})\ ,\
\vec f_{23}(s) = (0\frac{1}{2}\frac{1}{2}\frac{1}{2})\ ,\
\vec f_{23}(t) = (\frac{1}{2}000)
\nonumber \\
\vec f_{23}(b) & = & (\frac{1}{2}\frac{1}{2}00)\ ,\
\vec f_{23}(\mu) = (\frac{1}{2}\frac{1}{2}0\frac{1}{2})\ ,\
\vec f_{23}(\tau) = (0\frac{1}{2}0\frac{1}{2})
\label{Z4ass}
\end {eqnarray}
where $\vec{f_{23}}(\phi)$ is the corresponding $\vec{f_{23}}$ (see
eq.(\ref{ac41})) for the $h_\phi$ coupling, remarkably good fits
can be found. Eq.(\ref{Z4ass}) is satisfied (up to spurious lattice
vectors) by the following assignment of physical fields to
$\{\theta^k,f\}$ sectors
\begin {eqnarray}
Q_c & : & (\frac{1}{2}\frac{1}{2}00),\;\;Q_t:(0000),\;\;
c^c:(\frac{1}{2}\frac{1}{2}\frac{1}{2}\frac{1}{2}),\;\;
s^c:(0\frac{1}{2}\frac{1}{2}\frac{1}{2})
\nonumber \\
b^c & : & (\frac{1}{2}\frac{1}{2}00),\;\;t^c:(0\frac{1}{2}00),\;\;
L_\mu:(00\frac{1}{2}\frac{1}{2}),\;\;
L_\tau:(\frac{1}{2}\frac{1}{2}\frac{1}{2}\frac{1}{2})
\nonumber \\
\mu^c & : & (\frac{1}{2}\frac{1}{2}0\frac{1}{2}),\;\;\tau^c:(0\frac{1}{2}
0\frac{1}{2}),\;\;
H_1:(0000),\;\;
H_2:(\frac{1}{2}\frac{1}{2}00)
\label{Z4ass2}
\end {eqnarray}
where the $\phi^c$ fields are understood to belong to the $\theta^2$ sector
and the rest to the $\theta$ one\footnote{There are other possible
assignments consistent with (\ref{Z4ass}).}.
The values for the deformation parameters (in string units) and $\nu_1$
for an illustrative fit are
\begin {eqnarray}
R_1 &=& 13.280\;,\;\;R_3 = 15.077\;,\;\;\alpha_{13}=-0.2395
\nonumber \\
\alpha_{14} &=&0\;,\;\;\nu_1=71.8\ GeV
\label{Z4fit}
\end {eqnarray}
and the corresponding fermion masses are shown in the third row of Table 2.
(Oscillations around these values with the subsequent variations of the
fermions masses are of course possible.)
This result is rather remarkable, specially when one notices that
the number of free parameters is lower than the number of physical
masses fitted. In some sense the experimental masses, which are low--energy
quantities, are selecting a particular assignment of the physical fields
to fixed points and the values of the deformation parameters that
define the size and shape of the compactified space (e.g. the "preferred"
$\theta_{14}$ angle is the cartesian one). We find this
quite encouraging. Notice also that the numbers of eq.(\ref{Z4fit})
are quite sensible for a compactified space. The hierarchy of masses
which emerges from them has to do with the exponential dependence
of the Yukawa couplings (see eq.(\ref{ac41})).

On the technical side, let us comment that the fit has been performed
with the help of a MINUIT program, choosing for the minimization
function the total $\chi^2$. The major obstacle we have found was
to control the convergence of the Jacobi $\vartheta$ function of
eq.(\ref{ac41}), particularly when the $\alpha_{13}$, $\alpha_{14}$
parameters are close to the boundary of their definition range.
This requires to sum up to 10000 terms of the series: it is
by no means significant to keep only a few terms. Finally, we
have increased the usual experimental errors of $m_\mu$ and
$m_\tau$ up to $1\%$ to incorporate, to some extent, the errors
attributable to the calculation. The "experimental" error of $m_b$ was
conservatively set at $\Delta m_b=10\%$.

Let us now comment the results for the other $Z_N$ orbifolds.
It turns out that, besides $Z_4$, the only ones that can work
are the $Z_3$ and the $Z_6$--I. Our best fits for them are
shown in the second and fourth rows of Table 2.
For the $Z_6$--I one, all the couplings considered were of the
$\theta^2\theta^2 \theta^2$ type. Consequently, all the physical
fields are understood to belong to the $\theta^2$ sector in this case.
On the other hand, in the $Z_3$ orbifold there is a unique $\theta$ sector.
The corresponding
assignments, given in the respective lattice basis, are
\begin {eqnarray}
&f_{23}(c)&  \left\{\begin{array}{cc}
Z_3 & :  (00\frac{1}{3}\frac{2}{3}\frac{1}{3}\frac{2}{3}) \\
Z_6 & :  (0\frac{1}{3}0\frac{1}{3}00)
\end{array}
\right.
,\;\;
f_{23}(s)    \left\{\begin{array}{cc}
Z_3 & :  (\frac{1}{3}\frac{2}{3}\frac{2}{3}\frac{1}{3}\frac{1}{3}
\frac{2}{3}) \\
Z_6 & :  (0\frac{1}{3}00\frac{1}{3}\frac{2}{3})
\end{array}
\right.
,\;\;
f_{23}(t)  \left\{\begin{array}{cc}
Z_3 & :  (0000\frac{1}{3}\frac{2}{3}) \\
Z_6 & :  (0000\frac{1}{3}\frac{2}{3})
\end{array}
\right.
\nonumber \\
&f_{23}(b)& \left\{\begin{array}{cc}
Z_3 & :  (\frac{1}{3}\frac{2}{3}\frac{2}{3}\frac{1}{3}00) \\
Z_6 & :  (0\frac{1}{3}0000)
\end{array}
\right.
,\;\;
f_{23}(\mu)  \left\{\begin{array}{cc}
Z_3 & :  (\frac{1}{3}\frac{2}{3}\frac{1}{3}\frac{2}{3}\frac{1}{3}
\frac{2}{3}) \\
Z_6 & :  (000\frac{1}{3}\frac{1}{3}\frac{2}{3})
\end{array}
\right.
,\;\;
f_{23}(\tau)   \left\{\begin{array}{cc}
Z_3 & :  (\frac{1}{3}\frac{2}{3}\frac{1}{3}\frac{2}{3}00) \\
Z_6 & :  (000\frac{1}{3}00)
\end{array}
\right.
\label{Z36ass}
\end {eqnarray}
where, again, $f_{23}(\phi)$ is the difference between two of the fixed
points involved in the $h_\phi$ coupling. The corresponding values for
the nine deformation parameters that the $Z_3$ orbifold possesses (see
Table 1 and ref.[6]) and for $\nu_1$ are
\begin {eqnarray}
R_1 &=&13.039 \;,\;\;R_3 = 30.460\;,\;\;R_5=13.076
\nonumber \\
\alpha_{13} &=&-0.6055\;,\;\alpha_{14} = -0.2395\;,\;\;\nu_1=39.3\ GeV
\nonumber \\
\alpha_{15} &=& \;\;\alpha_{16} = \;\;\alpha_{35} = \;\;\alpha_{36}
= 0\; (fixed)
\label{Z3fit}
\end {eqnarray}
Notice that four of them have not been used in the fit. This has been done
to improve the convergence of the MINUIT program. Clearly, a better
fit could be obtained once these four parameters are also considered.
Similarly, the values for the five deformation parameters of the $Z_6$--I
orbifold and $\nu_1$ are
\begin {eqnarray}
R_1 &=& 18.349\;,\;\;R_3 = 17.588\;,\;\;R_5=13.073
\nonumber \\
\alpha_{13} &=& -0.3438\;,\;\alpha_{14} = 0.2978\;,\;\;\nu_1=70\ GeV
\label{Z6fit}
\end {eqnarray}

For the rest of the orbifolds, after an exhaustive exploration,
we have not found sensible fits. This should not be surprising
since they have a smaller number of effective deformation parameters,
see Table 1. This consideration makes the $Z_4$ case the most
remarkable one. Just for completeness we have given in Table 2
our best fits for these orbifolds. It is worth noticing that
the $Z_7$ orbifold (which has only four different couplings and three
deformation parameters) works very acceptably for all fermions masses,
except for the strange one, which on the other hand has a large experimental
uncertainty.

Let us finally give the values of the moduli $T_i$ corresponding
to eqs.(\ref{Z4fit},\ref{Z3fit},\ref{Z6fit}). As usual, we define
the normalization of $T_i$ in such a way that, under a duality
transformation, they transform as $T_i\rightarrow 1/T_i$ [29]. This
implies Re $T_{i}= \alpha R_{i}^2$ with $\alpha=\frac{\sqrt{3}}{16\pi^2}$
for $Z_3$ and $Z_6$, and $\alpha=\frac{\sqrt{2}}{8\pi^2}$ for
$Z_4$.\footnote{Notice that these values of $\alpha$ are
consistent with the definition of our lattice. This
is the usual one, coinciding with that of the first paper
of ref.[2], while in the second paper of the same reference
the lattice is redefined as $e_i\rightarrow 2\pi e_i$.
Of course the values of $T_i$ are independent on these
redefinitions.} Hence, $T_1=1.86$, $T_3=10.17$,
$T_5=1.87$ (for $Z_3$); $T_1=3.16$, $T_3=4.07$ (for $Z_4$) and
$T_1=3.69$, $T_3=3.39$, $T_5=1.87$ (for $Z_6$).

\section{Summary and Conclusions}

We have explored the capability of string theories to reproduce
the observed pattern of quark and lepton masses and mixing angles.
We have focused our attention on orbifold constructions since,
apart from their phenomenological merits, there is at present a
good knowledge of the theoretical Yukawa couplings in these scenarios.
A first conclusion is that, due to stringy selection rules, there
is no possibility of fitting the Kobayashi--Maskawa parameters
at the renormalizable level. This is so even for non--prime orbifolds,
where the mass matrices are allowed to have a non--diagonal structure.
It is, however, argued that (model--dependent) non--renormalizable couplings
might well do this job at the same time as they account for the masses of
the first generation (which should come from off--diagonal entries
in the mass matrices).
On the other hand, non--renormalizable couplings are
too suppressed to adequately fit the fermion masses of the second
and third generations ($m_\mu$, $m_\tau$, $m_c$, etc.), which, in consequence,
should be accounted for by renormalizable ones (this is still extremely
restrictive). We then examined this issue.

As a first step, a renormalization group running of the Yukawa couplings
between the string scale ($M_{Str}$) and the low--energy scale ($1\ GeV$)
has to be performed. This running is slightly different from the
ordinary GUT one since the gauge couplings are not unified at $M_{Str}$
due to string threshold corrections. This modifies, for example,
the traditional $m_b/m_\tau$ relation at low energy, although not
substantially. The magnitude of the orbifold Yukawa couplings
depends on the values of the so--called deformation parameters,
which describe the size and shape of the compactified space. A
systematic exploration allows us to check whether there is a choice
of these deformation parameters for which the physical fermion masses
are properly fitted. Not all the $Z_N$ orbifolds are here on a
same footing. It turns out that this fit is possible only for the
$Z_3$, $Z_4$ and $Z_6$--I orbifolds. Besides these, the $Z_7$ orbifold
is able to fit all the fermion masses except the strange one, which
on the other hand has a large experimental uncertainty.

The corresponding values of the
deformation parameters are quite reasonable (they correspond to
values for the moduli $T_i=O(1)$). The case of the $Z_4$ orbifold
is specially remarkable since the number of free parameters is
lower than the number of physical masses fitted. In some sense
the experimental masses, which are low--energy quantities, are selecting
a particular size and shape of the compactified space. We find this
quite encouraging. It should be stressed, however, that this only
shows the compatibility of certain string schemes with the
low--energy measurements, although this is certainly non--trivial.
The rest of $Z_N$ orbifolds, however, are rather hopeless
and should be discarded on these grounds. Finally, let us remark that
all these results have been obtained under the assumptions explained
in the Introduction, in particular within a minimal $SU(3)\times
SU(2)\times U(1)_Y$ scenario.


\newpage

\textwidth=18cm
\textheight=22cm
\oddsidemargin=-1cm
\evensidemargin=0cm
\topmargin=0cm
\parindent=1cm

\begin{table}
\underline{\bf TABLE 1}
$\begin{array}{|c|c|c|c|c|c|c|c|c|} \hline
Orb. & Twist\;\theta  & Lattice & \#DP & Coupling  & \#AC
&\#EDP  &\#DCR  & \#DCD \\ \hline
Z_3  & (1,1,-2)/3    & SU(3)^3 & 9 & \theta\theta\theta    & 729 &
9 & 4 & 14 \\ \hline
Z_4  & (1,1,-2)/4    & SU(4)^2 & 7 & \theta\theta\theta^2    & 160 &
4 & 6 & 10 \\
  &     & SO(4)^3 & 7 & \theta\theta\theta^2    & 160 &
4 & 6 & 8 \\ \hline
Z_6-{\rm I} & (1,1,-2)/6 & G_2^2\times SU(3) & 5 & \theta\theta^2\theta^3 & 90
&
4 & 10 & 30 \\
      &  &  & & \theta^2\theta^2\theta^2 & 369 &
5 & 8 & 12 \\ \hline
Z_6-{\rm II} & (1,2,-3)/6 & SU(6)\times SU(2) & 5 & \theta\theta^2\theta^3
& 48 & 1 & 4 & 4 \\
      &  &  & & \theta\theta\theta^4 & 72 &
2 & 4 & 4 \\ \hline
Z_7 & (1,2,-3)/7 & SU(7) & 3 & \theta\theta^2\theta^4
& 49 & 3 & 2 & 4 \\ \hline
Z_8-{\rm I} & (1,2,-3)/8 & SO(5)\times SO(9) & 3 & \theta^2\theta^2\theta^4
& 80 & 2 & 8 & 8 \\
      &  &  & & \theta\theta^2\theta^5 & 40 &
3 & 8 & 9 \\ \hline
Z_8-{\rm II} & (1,3,-4)/8 & SO(4)\times SO(8) & 5 & \theta\theta\theta^6
& 24 & 2 & 3 & 3 \\
                                                &  &  & &
\theta^2\theta^3\theta^3 & " &
" & " & " \\
      &  &  & & \theta\theta^3\theta^4 & 48 &
2 & 6 & 6 \\ \hline
Z_{12}-{\rm I} & (1,4,-5)/12 & SU(3)\times F_4 & 3 & \theta\theta^2\theta^9
& 6 & 2 & 2 & 2 \\
      &  &  & & \theta^2\theta^3\theta^7 & " &
" & " & " \\
      &  &  & & \theta\theta^4\theta^7 & 27 &
3 & 4 & 6 \\
      &  &  & & \theta^2\theta^4\theta^6 & 36 &
2 & 7 & 12 \\
      &  &  & & \theta^4\theta^4\theta^4 & 135 &
3 & 8 & 12 \\ \hline
Z_{12}-{\rm II} & (1,5,-6)/12 & SO(4)\times F_4 & 5 & \theta\theta\theta^{10}
& 4 & 2 & 1 & 1 \\
      &  &  & & \theta^2\theta^5\theta^5 & " &
" & " & " \\
      &  &  & & \theta\theta^3\theta^8 & 24 &
2 & 6 & 6 \\
      &  &  & & \theta^3\theta^4\theta^5 & 24 &
2 & 6 & 6 \\
      &  &  & & \theta^3\theta^3\theta^6 & 40 &
2 & 6 & 8 \\
      &  &  & & \theta\theta^5\theta^6 & 16 &
2 & 3 & 4 \\ \hline
\end{array}$
\caption{Characteristics of twisted Yukawa couplings for $Z_n$
orbifolds. The twist $\theta$ is
specified by the three $c_i$ parameters (one for each complex
plane rotation) appearing in $\theta=\exp (\sum c_i J_i)$.
$\#DP\equiv$ No. of deformation parameters, $\#AC\equiv$ No. of
allowed couplings, $\#EDP\equiv$ No. of effective deformation
parameters, $\#DCR\equiv$ No. of different Yukawa couplings for
the non--deformed (rigid) orbifold, $\#DCD\equiv$ No. of different
Yukawa couplings when deformations are considered. Quotation marks
denote equivalent couplings.}

\end{table}

\newpage

\begin{table}
\underline{\bf TABLE 2}
$\begin{array}{|c|c|c|c|c|c|c|c|} \hline
{} & m_\mu  & m_\tau & m_s & m_c  & m_b
& m_t &\chi^2_{tot} \\ \hline
Exp.  & 0.1056 & 1.784 & 0.199 & 1.35 & 5 & 130 & -  \\ \hline
Z_3   & 0.1055 & 1.786 & 0.252 & 1.35 & 4.1 & 125 & 6.7  \\ \hline
Z_4   & 0.1062 & 1.774 & 0.173 & 1.35 & 4.34 & 104 & 3.86  \\ \hline
Z_6-{\rm I}   & 0.1056 & 1.785 & 0.252 & 1.35 & 4.04 & 122 & 6.8  \\ \hline
Z_6-{\rm II}  & 1.13 & 64 & 5.1 & 17 & 8.3 & 173 & 3\times 10^8  \\ \hline
Z_7   & 0.104 & 1.783 & 0.466 & 1.35 & 5.2 & 133 & 81  \\ \hline
Z_8-{\rm I}   & 0.087 & 2.00 & 0.280 & 2.17 & 6.3 & 36 & 705  \\ \hline
Z_8-{\rm II}  & 0.1058 & 1.82 & 0.009 & 0.89 & 2.47 & 172 & 158  \\ \hline
Z_{12}-{\rm I}   & 0.103 & 1.83 & 0.107 & 1.40 & 11 & 88 & 177  \\ \hline
Z_{12}-{\rm II}  & 0.036 & 2.15 & 0.045 & 1.0 & 30 & 24 & 7236  \\ \hline
\end{array}$
\caption{Fits for each $Z_N$ orbifold of the first and second
generation fermion masses and total $\chi^2$. The masses,
given in $GeV$, are to be understood at the $1\ GeV$ scale, except
the top mass, which is at the $M_Z$ scale.
The first row corresponds to the present central
experimental values (errors are not shown). For the top mass, the
recent estimations based on the size of the electroweak radiative
corrections was considered. Only the $Z_3$, $Z_4$ and $Z_6$--I
orbifolds are compatible with the experiment.}

\end{table}

\vspace{3cm}
\noindent {\bf FIGURE CAPTION}

\vspace{2cm}
\noindent {\bf FIG.1}:  Values of $h_c,h_s,h_b,h_t,h_\mu,h_\tau$
at $M_{Str}$ versus $\nu_2=<H_2>$ giving the measured values of
the corresponding fermion masses. Errors are not included.

\end{document}